\documentclass[prl,aps,showpacs,twocolumn,floatfix]{revtex4-1}
\usepackage{graphicx}
\usepackage{bm}
\usepackage{amsmath}
\usepackage{amssymb}

\begin{document}

\title{Concavity Theorems for Energy Surfaces}

\author{B. G. Giraud}
\affiliation{Institut de Physique Th\'eorique,
Centre d'Etudes Saclay, 91190 Gif-sur-Yvette, France}
\email{bertrand.giraud@cea.fr}
\author{S. Karataglidis}
\affiliation{Department of Physics, University of Johannesburg,
P. O. Box 524, Auckland Park, 2006, South Africa}
\email{stevenka@uj.ac.za}

\begin{abstract}
Concavity properties prevent the existence of significant landscapes in
energy surfaces obtained by \textit{strict} constrained energy minimizations. 
The inherent contradiction is due to fluctuations of collective coordinates. 
A solution to those fluctuations is given.
\end{abstract}
\date{\today}
\maketitle

The concept of collective coordinates \cite{BohMot} has been of
central importance in atomic, molecular, and nuclear physics. Its goal has
been to generate models involving far less degrees of freedom than the true
number, $3 A$, as needed for a microscopic description of a system of $A$
particles.  Often, its dynamics can be compressed into slow motions of a few
collective degrees of freedom $B$, while the other, faster, degrees can be
averaged out. Also, for identical particles at least, such collective
degrees can be one-body operators,
$B=\sum_{i=1}^A \beta(\mathbf{r}_i,\mathbf{p}_i,\sigma_i,\tau_i),$ where 
$\mathbf{r}_i, \mathbf{p}_i, \sigma_i, \tau_i$ refer to the position,
momentum, spin, and if necessary isospin, respectively, of particle $i$.
The summation over $i$ provides, intuitively at least, a motivation 
for more inertia in $B$ than in the individual degrees $\beta_i$.

The concept of energy surfaces \cite{NixSwi} has been as important.
Given a ``coordinate-like'' collective operator $B$ and its expectation value 
$b \equiv \left\langle B \right\rangle$, most collective models (in a theory of
nuclear fission or fusion, for example) accept that there  exists an energy
function, $e(b)$, and an inertia parameter, $\mu(b),$ that drive the
collective dynamics. Keywords such as ``saddles'', ``barriers'', etc., flourish
\cite{NixSwi,FQVVK}.

Simultaneously, it is often assumed that the function, $e(b),$ results
from an energy \textit{minimization} under constraint. Namely, while
the system evolves through various values of $b$, it is believed to tune its
energy to achieve a (local) minimum. This aspect of finding $e(b)$
is central to many fields of physics. To illustrate, consider a Hamiltonian,
$H=\sum_i T_i + \sum_{i<j} V_{ij},$ where $T$ and $V$ denote the
usual kinetic and interaction operators. Given a trial set of density
operators, $\mathcal{D}$, in many-body space, normalized by
$\text{Tr} \mathcal{D}=1$, a prescription for $e(b)$ often reads,
\begin{equation}
e(b) = \inf_{\mathcal{D}\Rightarrow b} \text{Tr} \left\{ H \mathcal{D}
\right\},
\end{equation}
where $\text{Tr}$ is a trace in the many-body space for the $A$ particles.
The constraint, $\mathcal{D}\Rightarrow b,$ enforces $\text{Tr} \left\{ B
\mathcal{D} \right\}=b.$

There are theories which do not use, \textit{a priori}, an axiom of energy
minimization for the ``fast'' degrees. Time-dependent Hartree-Fock (HF)
\cite{Bonche} trajectories, generalizations with pairing, adiabatic versions
\cite{Villars}, often show collective motions. Equations of motion
\cite{Klein} and/or a maximum decoupling \cite{Matsuyanagi} of 
``longitudinal'' from ``transverse''  degrees, have also shown significant
successes in the search for collective degrees, at the cost, however, of
imposing a one-body nature of both collective coordinates and 
momenta and accepting state-dependence of these  operators. Such approaches
define an energy surface once trajectories of wave functions in many-body
spaces have been calculated. But they are not the subject of the present
analysis.  Herein, we focus on fixed operators constraining strict energy
minimizations within a fixed basis for single-particle and many-body states.

Ideally, to properly define a function $e(b)$ of the collective coordinate,
one should first diagonalize $B$ within the space provided by the many-body
states available for calculations. The resulting spectrum of $B$ is assumed
to be continuous, or at least have a high density for that
chosen trial space. Then, for each eigenvalue, $b$, one should find the ground
state eigenvalue, $e(b)$, of the projection of $H$ into that eigensubspace
labeled by $b$.

In practice, however, one traditionally settles for a diagonalization of the
constrained operator, $\mathcal{H} \equiv H - \lambda B$, where $\lambda$ is
a Lagrange multiplier,  or at least for a minimization of 
$\left\langle \mathcal{H} \right\rangle.$  Concomitantly, $B$ is assumed to 
have both upper and lower bounds, or, at least, that the constrained
Hamiltonian, $\mathcal{H},$ always has a ground state.
This returns the ``free energy'', 
$\varepsilon(\lambda) \equiv \left\langle \mathcal{H} \right\rangle$. The
label $b$ is no longer an eigenvalue but just an average value,
$b = \left\langle B \right\rangle$. A standard Legendre transform of $\langle H
\rangle$ then yields
the ``energy surface'', $e(b)$. Well known properties of this process are,
$d\varepsilon/d\lambda=-\text{Tr}\{ B\mathcal{D} \} = -b$, and, $de/db=\lambda$.
However we show in this work that constrained variation in a \textit{quantum}
system without additional precautions can raise at least two problems, namely:
i) there is a link between strict minimization and curvature
properties of $e(b)$; and, moreover, ii) the usual interpretation of the
parameter $b$ as a well-defined coordinate for a collective model can be
negated by non-negligible fluctuations.\\

\textit{Theorem linking strict minimization and concavity:}

Before proving the theorem, we must recall that, with Hartree-Fock (HF) and
Hartree-Bogoliubov (HB) approximations, both concave and
convex branches were obtained for $e(b)$ by replacing the constraint term,
$-\lambda \left\langle B \right\rangle$, in 
$\left\langle \mathcal{H} \right\rangle$
by either $-\lambda' \left\langle B \right\rangle + \mu \left\langle B 
\right\rangle^2/2,$
\cite{GLW}, or $-C \left( \left\langle B \right\rangle - \mu \right)^2/2$,
\cite{FQVVK}, with
adjustable values of $\lambda'$, $C$, and $\mu$. However, both methods, while
stabilizing the numerical procedure, amount to use an effective Lagrange 
multiplier, namely 
$\lambda_{\text{eff}} = \lambda - \left\langle B \right\rangle \mu$ and
$\lambda_{\text{eff}} = 
C \left( \left\langle B \right\rangle - \mu \right)$, respectively. We shall,
therefore, stick to the generic form, $H - \lambda B$, in the following.

Consider a solution branch $\mathcal{D}(\lambda)$, expanding 
up to second order, and assuming that the manifold of 
solutions is suitably analytic,
\begin{equation}
\mathcal{D}(\lambda+d \lambda) = \mathcal{D}(\lambda)+d\lambda
 (d\mathcal{D}/d\lambda) + (d\lambda^2/2) (d^2\mathcal{D}/d\lambda^2).
\end{equation}
The stationarity and minimality of 
$\text{Tr}\left\{ \mathcal{H} \mathcal{D} \right\}$ with respect to any
variation of $\mathcal{D},$ and in particular w.r.t. that variation,
$\mathcal{D}(\lambda+d\lambda)-\mathcal{D}(\lambda),$ induce,
\begin{align}
\text{Tr} \left\{ \mathcal{H} d\mathcal{D}/d \lambda\right\} & = 0, \nonumber
\\ 
\text{Tr} \left\{ \mathcal{H}d^2\mathcal{D}/d \lambda^2 \right \} & \ge 0.
\label{gensolcond}
\end{align}
The free energy $\varepsilon$ is also stationary for 
$\mathcal{D}(\lambda + d\lambda)$, but the Hamiltonian is now, 
$\mathcal{H}(\lambda) - B d\lambda$, and the derivative of the state is,
$d\mathcal{D}/d\lambda + d\lambda (d^2\mathcal{D}/d\lambda^2) +
\mathcal{O}(d\lambda^2)$, hence,
\begin{equation}
\text{Tr} \left\{ \left( \mathcal{H} - Bd\lambda\right) 
\left[ d\mathcal{D}/d\lambda + 
d\lambda (d^2\mathcal{D}/d\lambda^2)+\mathcal{O}(d\lambda^2) \right] \right\} = 0.
\label{genstatio}
\end{equation}
The zeroth order of this, Eq.~(\ref{genstatio}), is, 
$\text{Tr} \left\{\mathcal{H} d\mathcal{D}/d\lambda \right\}$. It vanishes,
because of the first of Eqs.~(\ref{gensolcond}). The first order, once
divided by $d \lambda$, gives,
\begin{equation}
- \text{Tr} \left\{B d\mathcal{D}/d\lambda \right\} =
- \text{Tr} \left\{\mathcal{H} d^2\mathcal{D}/d\lambda^2 \right\}.
\label{trick}
\end{equation}
The left-hand side of Eq.~(\ref{trick}) is nothing but the the second 
derivative, $d^2 \varepsilon/d\lambda^2$. The right-hand side is
semi-negative-definite, because of the second of Eqs.~(\ref{gensolcond}).
Hence, the plot of $\varepsilon(\lambda)$ is a convex curve and the plot
of its Legendre transform, $e(b)$, is concave. (Other authors may have the
opposite sign convention of the second derivative to define concavity versus
convexity.) With our sign convention \cite{BG}, strict minimization
necessarily induces concavity, and any convex branch means that the 
``fast'' degrees of freedom are not in a minimal energy.

It is important to note that this proof does not assume any specification of 
$\mathcal{D}(\lambda)$, whether it is constructed either from exact or
approximate eigenstates of $\mathcal{H}$. Therefore strict minimization can
\textit{only} return \textit{concave} functions $e(b)$. Maxima are impossible.
In the generalization where several collective operators 
$B_1, \dots, B_N$, are involved, concavity stills holds, so saddles
are also excluded. Hence, only an absolute minimum is possible. (However, we
shall show below how to overcome the paradox: by \textit{keeping
constant the fluctuations} of the collective coordinate(s), one can deviate 
from concavity, and more important, validate a constant quality of the 
representaion provided by branches $\mathcal{D}(\lambda)$.)\\

\textit{Same theorem, for diagonalizations:}

Let  $\psi(\lambda)$ be the ground state of $\mathcal{H}$. (For the sake of
simplicity, we assume that there is no degeneracy.) The corresponding
eigenvalue, $\varepsilon(\lambda)$, is stationary with respect to variations
of $\psi$, among which is the ``online'' variation,
$d\lambda\, (d\psi/d\lambda)$, leading to the well-known first derivative,
$d\varepsilon/d\lambda = -b \equiv - \left\langle \psi \left| B \right|\psi 
\right\rangle$.
Consider the projectors
$P = \left| \psi \right\rangle \left\langle \psi \right|$ and $Q = 1-P.$
Brillouin-Wigner theory yields the first derivative of $\psi$, \textit{viz.}
\begin{equation}
\frac{d\left|\psi \right\rangle}{d\lambda} =
- \frac{Q}{\varepsilon-Q\mathcal{H}Q} B \left| \psi \right\rangle.
\end{equation}
This provides the second derivative of $\varepsilon,$
\begin{equation}
-\frac{d b}{d \lambda} \equiv - \frac{d}{d \lambda} \left\langle \psi \left|
 B \right| \psi \right\rangle = 2 \left\langle \psi \left| B 
\frac{Q}{\varepsilon-Q\mathcal{H}Q} B \right| \psi \right\rangle.
\label{negatdefin}
\end{equation}
Since the operator $(\varepsilon-Q\mathcal{H}Q)$ is clearly
negative-definite, the eigenvalue, $\varepsilon$, is a convex function of
$\lambda$. It is trivial to prove that the same convexity holds for the
ground state eigenvalue $\varepsilon(\lambda_1, \dots,\lambda_N)$ if several
constraints, $B_1, \dots, B_N$, are used. If, moreover, a temperature 
$T$ is introduced, the thermal state, 
$\mathcal{D}=\exp\left[-\mathcal{H}/T\right] / 
\text{Tr}\exp\left[- \mathcal{H}/T\right],$
replaces the ground state projector, 
$\left| \psi(\lambda_1,\dots,\lambda_N) \right\rangle
\left\langle \psi(\lambda_1,\dots,\lambda_N) \right|,$ and
the free energy, $\varepsilon(\lambda_1, \dots,\lambda_N;T)$,
also contains the entropy contribution, $-T S,$ where
$S=-\text{Tr} \left\{ \mathcal{D} \ln \mathcal{D} \right\}$. A proof of the
convexity of the exact $\varepsilon(\lambda_1, \dots,\lambda_N;T)$ is also
easy \cite{Balian}.

At $T=0,$ the usual Legendre transform expresses the energy,
$e \equiv \left\langle \psi \left| H \right| \psi \right\rangle$, in terms of
the constraint value(s) rather than the Lagrange multiplier(s). For
simplicity, consider one constraint only; the generalization to $N>1$ is
easy. Since $e \equiv \varepsilon + \lambda b$, then
$d e/d b = \lambda$, a familiar result for conjugate variables.
Furthermore, the second derivative, $d^2 e/d b^2,$ reads, 
$d\lambda/db = 1/(db/d\lambda)$. From Eq.~(\ref{negatdefin}), the derivative,
$db/d\lambda$, is positive-definite. Accordingly, $e$ is a concave function
of $b$. Now, if $T > 0$, the Legendre transform instead generates a reduced
free energy, $\eta \equiv (e - TS)$, a concave function of the constraint
value(s). An additional Legendre transform  returns $e$ alone, as a concave
function of the constraint(s) and $S$.

Let $b_-$ and $b_+$ be the lowest and highest eigenvalues of $B.$
When $\lambda$ runs from $-\infty$ to $+\infty,$  then $b$
spans the interval, $[b_-,b_+]$. There is no room for a junction with convex 
branches under technical modifications as used by \cite{GLW,FQVVK}.
For every exact diagonalization of $\mathcal{H}$, or exact partition function,
concavity sets a one-to-one mapping between $b$ in this interval and $\lambda$.
More generally, with exact calculations, there is a one-to-one mapping
between the set of Lagrange multipliers,
$\left\{\lambda_1, \dots,\lambda_N \right\},$
and that of obtained values, $\left\{ b_1, \dots, b_N \right\}$, of the
constraints. Concavity, \textit{in the whole obtained domain of constraint
values}, imposes a poor landscape: there is one valley only.

We tested this surprising result with several dozens of numerical cases, 
where we used random matrices for $H$ and $B,$ with various dimensions. As
an obvious precaution, we eliminated those very rare cases where both $H$ and
$B$ turned out, by chance, to be block matrices with the same block structure;
such cases allow level crossings and degeneracies. Then every remaining 
situation, without any exception, confirmed the concavity theorem.

\begin{figure}
\scalebox{0.68}{\includegraphics*{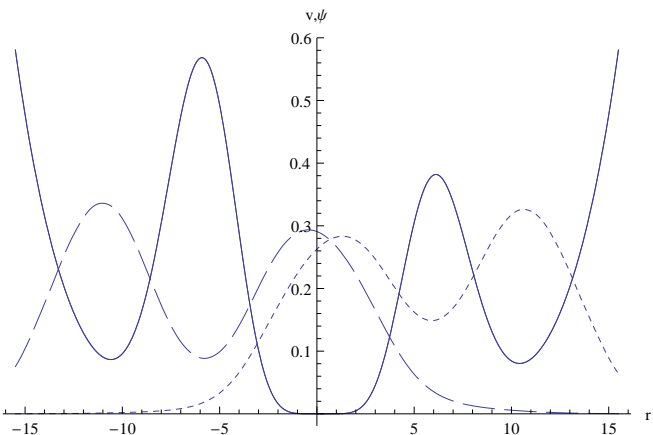}}
\caption{\label{figure1} Toy model, 1-D potential, full curve. 
Tunnel effect of constrained eigenstate under left barrier, dashed curve. 
Same effect under right barrier, dotted curve.}
\end{figure}

\begin{figure}
\scalebox{0.68}{\includegraphics*{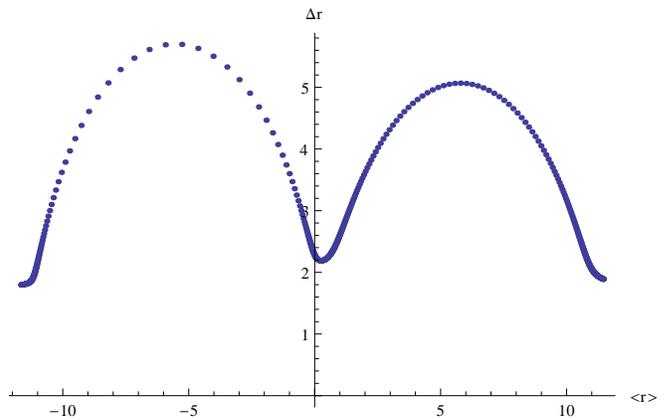}}
\caption{\label{figure2} The fluctuation, $\langle \psi | r^2 | \psi \rangle 
- \langle \psi | r | \psi \rangle^2,$ increases when the constrained
eigenstate delocalizes into two wave packets.}
\end{figure}

Such a paradox of energy concavity even in the presence of a 
``wavy'' potential can be understood from, for instance, Fig. \ref{figure1}. 
A toy, one-dimensional, Hamiltonian, $h = -d^2/dr^2+v(r)-\lambda r$, with a
double hump potential, $v=(r-1/5)^6/5000 \exp(-r^2/12) + (r/12)^6/8$, shown
as a full curve presented in Fig. \ref{figure1}, is
diagonalized in a subspace spanned by shifted Gaussians. While the eigenstate,
$\psi_{\lambda}(r),$ is made of a single wave-packet when the average value, 
$\left\langle r \right\rangle \equiv \left\langle \psi_{\lambda} \left| r 
\right| \psi_{\lambda} \right\rangle,$ sits near a minimum of $v$, a striking
tunnel effect occurs when $\left\langle r \right\rangle$ sits near a maximum
of $v$. There $\psi$ shows two connected packets, one at each side of the
barrier, inducing a lowering of the energy. Such a bimodal (even multimodal
in the extreme cases we tested) situation makes a very bad probing of the
barrier. When tunnel effects occur, fluctuations, 
$\Delta r = \sqrt{\left\langle r^2 \right\rangle - \left\langle r 
\right\rangle^2},$
are dramatically larger than when a unique packet sits in a valley. The
collective label, $\langle r \rangle$ in our case, is misleading. Although
$\mathcal{H}$ was exactly diagonalized, constrained variation generated a bad
quality representation of $v(r)$. Fig. \ref{figure2} illustrates how big the
error bar on $\left\langle r \right\rangle$ can become.

A trivial way to prevent fluctuations from arbitrarily varying is to 
introduce a double constraint via the square, $B^2$, of the initial constraint 
operator and adjust the second Lagrange multiplier so that the fluctuation, 
$\Delta b \equiv \sqrt{\left\langle B^2 \right\rangle - 
\left\langle B \right\rangle^2},$ remains small,
or at least, for a stable quality of the representation, reasonably constant.
Again with 1-D toy Hamiltonians of the form, 
$h'=-d^2/dr^2+v(r)-\lambda_1 r+\lambda_2 r^2,$
or, equivalently, $h'=-d^2/dr^2+v(r)+\lambda_2(r-\lambda_1)^2,$ we tuned 
$\lambda_2$ into a function $\lambda_2(\lambda_1)$ to enforce a unimodal 
situation  with $\Delta r$ kept constant when $\langle r \rangle$ evolves.
Concavity is then ``defeated''. A sharp probe of the barrier can be found. \\

\textit{Same theorem, for constrained HF calculations:}

Consider, now, energy surfaces obtained from approximations. Typically,
the minimization could be estimated from a HF or HB calculation,
at zero or finite $T$. Trial states used in such methods span a nonlinear
manifold; indeed, a sum of two determinants is usually not a
determinant. Let $\mathcal{D}(\lambda)$ denote one $A$-body
density operator where, within such nonlinear approximations, a minimum of 
$\text{Tr} \left\{\mathcal{H} \mathcal{D}\right\}$ or of
$\left(\text{Tr} \left\{\mathcal{H} \mathcal{D}\right\}-TS\right)$ is reached. 
Let $\varepsilon(\lambda)$ denote this minimum value. It may be degenerate,
but, in any case, it is stationary for arbitrary variations 
$\delta \mathcal{D}$ within the set of trial states. Accordingly, the first
derivative again reads,
$d \varepsilon/d \lambda = -b \equiv - \text{Tr} \left\{ B
 \mathcal{D}\right\}$. Then, if a Legendre transform holds, defining
$\eta \equiv \varepsilon + \lambda b$ in terms of $b$, the same argument that
was used for the exact case again yields, $d \eta/d b = \lambda$. With
$N$ constraints, the gradient of $\eta$ in the domain spanned by
$\left\{b_1,\dots,b_N \right\}$ is the vector $\left\{\lambda_1,\dots,\lambda_N \right\}$.

To discuss second derivatives, consider, for instance,
HF calculations, where $A$-body density operators are dyadics of determinants,
$\mathcal{D}= \left| \phi \right\rangle \left\langle \phi \right|$.
Norm-conserving variations of an HF solution, $\phi$, can be parametrized
as, $\left| \delta \phi \right\rangle = \exp(i X \delta \alpha ) \left| 
 \phi \right\rangle,$  with $X$ an arbitrary particle-hole Hermitian
operator, and $\delta \alpha$ an infinitesimal coefficient. Under such a
variation in the neighborhood of an HF solution, the first and second order
variations of the free energy,
$\varepsilon \equiv \text{Tr} \left\{ \mathcal{H} \exp(i X \delta \alpha ) 
\mathcal{D} \exp(-i X \delta \alpha ) \right\}$, read,
\begin{equation}
\delta \varepsilon=i \delta \alpha \text{Tr} \left\{ \left[\mathcal{H},X\right]
\mathcal{D} \right\} = 0,
\label{statio}
\end{equation}
and
\begin{equation}
\delta^2 \varepsilon=-\left(\delta \alpha^2/2\right) \text{Tr} 
\left\{ \left[\left[ \mathcal{H},X\right],X\right]\mathcal{D} \right\} \ge 0,
\label{mini}
\end{equation}
respectively. If $\mathcal{D}$ is an HF solution, the first order
vanishes $\forall X.$ Since only those solutions that give
minima are retained, the second order variation of $\varepsilon$ is 
semi-positive-definite, $\forall X$ again. Now, when $\mathcal{H}$ receives
the variation, $- B d \lambda$, there exists a particle-hole operator, $Y$, a
special value of $X$, that, with a coefficient $d\lambda$, accounts for
the modification of the solution. This reads
$\left| \Phi \right\rangle = \exp(i  Y d\lambda ) \left|\phi \right\rangle.$
Simultaneously,
those particle-hole operators that refer to this new Slater determinant 
$\Phi$ become $\mathcal{X}=\exp(i Y d\lambda) X \exp(-i Y d\lambda)$. 
The new energy is thus, 
\begin{equation}
\mathcal{E} = \text{Tr}\left\{\exp(-i Y d\lambda ) (\mathcal{H} - B d\lambda)
\exp(i Y d\lambda )\mathcal{D} \right\}. 
\end{equation}
The stationarity condition, Eq. (\ref{statio}), becomes,
\begin{align}
0 & = \text{Tr}\left\{\exp(-i Y d\lambda)\left[ (\mathcal{H} - Bd\lambda),
\mathcal{X}\right] \exp(i Yd\lambda) \mathcal{D}\right\} \nonumber \\
& = \text{Tr} \left\{\left[ \exp(-i Y d\lambda)(\mathcal{H} - B d \lambda)
\exp(i Yd\lambda) , X \right] \mathcal{D} \right\}.
\label{newstatio}
\end{align}
The zeroth order in $d\lambda$ of this, Eq.~(\ref{newstatio}), reads,
$\text{Tr}\left\{\left[ \mathcal{H}, X \right]\mathcal{D}\right\}$, and
vanishes $\forall X$, because of Eq.~(\ref{statio}). Then the first order in
$d\lambda$ gives, again $\forall X$,
\begin{equation}
\text{Tr} \left\{\left[ B,X \right]\mathcal{D}\right\} = 
i\text{Tr} \left\{\left[ \left[ \mathcal{H} , Y \right] , X \right]
\mathcal{D}\right\}.
\label{astuce}
\end{equation}
The second derivative is,
\begin{align}
d^2\varepsilon/d\lambda^2 & =-(d/d \lambda) \text{Tr} \left\{
\exp(-i Y d\lambda) B \exp(i Yd\lambda)\mathcal{D} \right\}\nonumber \\ 
&  = -i \text{Tr} \left\{ \left[ B,Y \right] \mathcal{D} \right\}.
\end{align}
Upon taking advantage of Eq.~(\ref{astuce}), for $Y$ as a special case
of $X,$ this becomes,
\begin{equation}
d^2\varepsilon/d\lambda^2 =
\text{Tr} \left\{\left[\left[ \mathcal{H},Y \right],Y\right ] \mathcal{D}
\right\},
\end{equation}
the right-hand side of which is semi-negative-definite, see Eq.~(\ref{mini}).
The solution branch obtained when $\lambda$ runs is, therefore, convex. Its
Legendre transform is concave.

The zoo of \textit{stationary} (not necessarily everywhere
minimizing) solutions of approximate methods may be rich enough to
accommodate inflections, where the curvature,
 $\text{Tr} \left\{ \mathcal{H}d^2\mathcal{D}/d\lambda^2 \right\}$,
can change sign. This makes a paradox: would non-linear approximations
generate a more flexible, physical tool than exact solutions?
``Phase transitions'', a somewhat incorrect wording for a finite system,
are sometimes advocated to accept continuing branches of energy minima
into metastable branches. But this definitely means dropping the axiom of
strict energy minimization to freeze fast degrees. This need for excited
solutions in constrained mean field calculations, namely two solutions for the
same value of $\lambda$ to describe both sides of an inflection point of the 
barrier, is well known and used. See in particular \cite{FQVVK}, where a
tangent parabola rather than a tangent straight line (that with slope 
$\lambda$) is used to explore the energy surface.

Besides the dropping of the ``fast degree minimization hypothesis'' one should
again wonder whether such mean field solutions, whether in concave or convex
branches, might be vitiated by large error bars for $b$. The following solvable
models give a preliminary answer.

Consider $N$ identical, 1-D fermions with Hamiltonian,
\begin{equation}
H=\sum_{i=1}^N p_i^2/(2m)+M \omega^2 R^2/2+\sum_{i>j=1}^N v_{ij},
\label{bareH}
\end{equation}
where $R=\sum_i r_i/N$ is the center-of-mass (c.m.) position, $p_i,r_i,m$
denote the single particle momentum, position and mass, respectively, of each
fermion, and $M=N m$ is the total mass. The c.m. momentum is, $P=\sum_i p_i.$
We use a system of units such that $\hbar=m=\omega=1,$ where $\omega$
denotes the frequency of the c.m. harmonic trap. The interaction, $v,$ is set
as Galilean invariant and so is the sum, $V=\sum_{i>j} v_{ij}.$ In the 
following, $v$ is taken as a spin and isospin independent and local force,
$v_{ij}=v(|r_i-r_j|).$

The collective operator we choose to constrain $H$ is a half sum of 
``inertia'' (mass weighted square radii), $B=\sum_{i=1}^N m r_i^2/2$. The
constrained Hamiltonian then reads,
\begin{equation}
{\cal H}=\sum_i p_i^2/(2m)+M \omega^2 R^2/2+\sum_{i>j} v_{ij}-\lambda
\sum_i m r_i^2/2.
\label{cnstHH}
\end{equation}

Let $\xi_1=r_2-r_1,$ $\xi_2=r_3-(r_1+r_2)/2,$ ..., 
$\xi_{N-1}=r_N-(r_1+r_2+\dots+r_{N-1})/(N-1)$ denote the usual Jacobi 
coordinates with $\Pi_{\alpha}$ and $\mu_{\alpha}$, $\alpha=1,\dots,(N-1)$,
the corresponding momenta and reduced masses. In this Jacobi
representation the constraint becomes, 
$B=M R^2/2+\sum_{\alpha} \mu_{\alpha} \xi_{\alpha}^2/2.$ Accordingly, the 
constrained Hamiltonian decouples as a sum of a cm harmonic oscillator,
\begin{equation}
\mathcal{H}_{\text{c.m.}} = \frac{P^2}{(2M)} + \frac{1}{2}M \Omega^2 R^2,
\; \; \; \Omega^2=\omega^2-\lambda,
\label{renortrap}
\end{equation}
provided  $\lambda<\omega^2$, and an internal operator,
\begin{equation}
\mathcal{H}_{\text{int}}=\sum_{\alpha} \left[ \Pi_{\alpha}^2/(2 \mu_{\alpha}) 
- \lambda \mu_{\alpha} \xi_{\alpha}^2/2 \right]+ V.
\label{intern}
\end{equation}
With the present power of computers and present experience with Faddeev and
Faddeev-Yakubovsky equations, this choice of $\mathcal{H}$ and $B$, with its
ability to decouple, provides soluble models with, typically,
$N=2,3,4$. (Decoupling also occurs if $B$ is a quadrupole operator.) 
Exact solutions can thus be compared with mean field approximations
and validate, or invalidate, the latter. In the present work, however, we are
not interested in the comparison, but just in properties of the mean 
field solutions as regards $B$.

For this, as long as $N$ does not exceed $4$, we assume that the
Pauli principle is taken care of by spins and isospins, understood in the
following, and that the space part of the mean field approximation is a
product, $\phi=\prod_i \varphi(r_i),$ of identical, real and positive parity
orbitals. The corresponding Hartree equation reads,
\begin{equation}
\left[\frac{p^2}{2m} + \Lambda r^2 + u(r) \right] \varphi(r)= \varepsilon_{sp}
\varphi(r),
\end{equation}
with $u(r)=(N-1) \int_{-\infty}^{\infty} v(r-s) [\varphi(s)]^2$ and 
$\Lambda=M \omega^2/(2 N^2)-\lambda m/2.$ The term,
$M \omega^2/(2 N^2),$ clearly comes from the c.m. trap. The same trap induces
a two-body operator, $M \omega^2 \sum_{i \ne j} r_i r_j/(2 N^2),$ which cannot
contribute to the Hartree potential, $u,$ since every dipole moment, 
$\left\langle r_j \right\rangle$, identically vanishes here.

Once $\varphi$ and $\varepsilon_{sp}$ are found, one obtains the free energy, 
$\varepsilon_{\text{Hart}}=\langle \mathcal{H} \rangle=N \varepsilon_{sp}-
N \left\langle 
\varphi \left| u \right| \varphi \right\rangle/2$, then the value of the
constraint, 
$\left\langle B \right\rangle_{\text{Hart}} = N m \left\langle \varphi 
\left| r^2 \right| \varphi \right\rangle/2,$ and
the square fluctuation, $(\Delta B)^2_{\text{Hart}} = N m^2 
\left( \left\langle \varphi \left|
 r^4 \right| \varphi \right\rangle - \left\langle \varphi \left| r^2 \right| 
\varphi \right\rangle^2 \right)/4.$
The physical energy, $e_{\text{Hart}}(b),$ in this Hartree approximation,
clearly obtains by adding $\lambda \langle B \rangle_{\text{Hart}}$ to 
$\varepsilon_{\text{Hart}}.$

\begin{figure}
\scalebox{0.68}{\includegraphics*{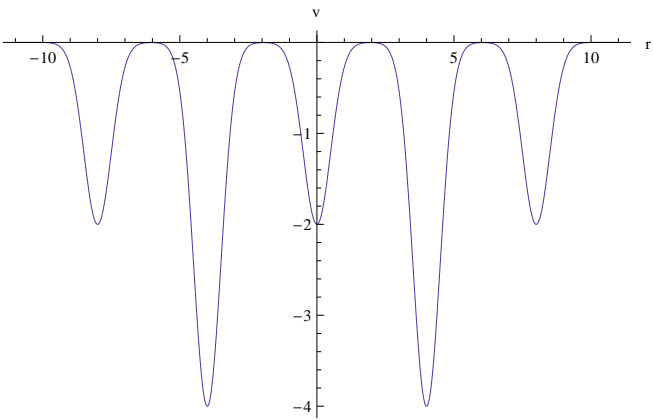}}
\caption{\label{figure3} An interaction giving multimodal Hartree solutions}
\end{figure}

We show now, among many cases we studied, Hartree results when 
$v_{ij}=-2 [\exp(-2 (r_i-r_j+8)^2)  +  \exp(-2 (r_i-r_j-8)^2) +
          2 \exp(-2 (r_i-r_j+4)^2) + 2 \exp(-2 (r_i-r_j-4)^2) + 
            \exp(-2 (r_i-r_j)^2)],$ see Fig. \ref{figure3}.

\begin{figure}
\scalebox{0.68}{\includegraphics*{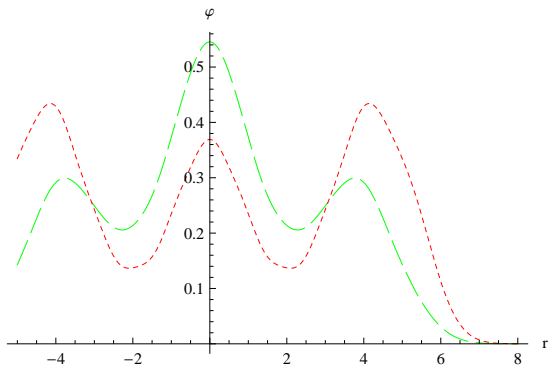}}
\caption{\label{figure4} Dashes, Hartree orbital $\varphi(r)$ if $N=2,$  
$\lambda=.47$ for the interaction shown in Fig. 3. Dots, the same for
$\lambda=.55.$}
\end{figure}

\begin{figure}
\scalebox{0.68}{\includegraphics*{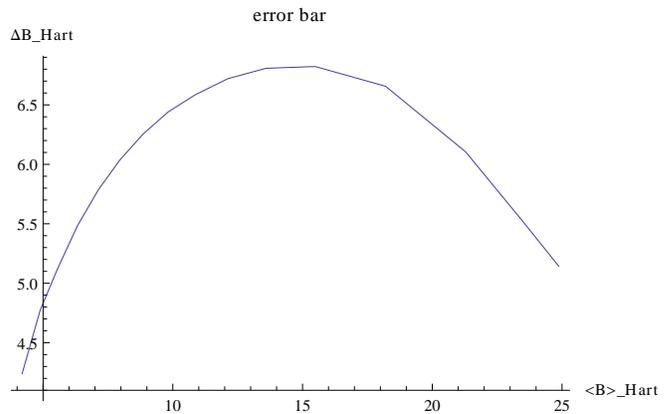}}
\caption{\label{figure5} Error bar $\Delta b$ as a function of 
$\langle B \rangle.$}
\end{figure}

The results shown in Figs. \ref{figure4}-\ref{figure5} correspond to $N=2$. 
Similar results hold with $N=3,4$. Orbitals $\varphi$ are expanded in the
first 11 even states of the standard harmonic oscillator. The only difference 
between the two orbitals shown in Fig. \ref{figure4} is the value of 
$\lambda$. Both orbitals, and many other ones, when $\lambda$ runs, show a 
bimodal structure, their left-hand-side and right-hand-side bumps being 
equivalent with respect to the even observable, $B \propto r^2$. Whether
$N=2$, $3$, or $4$, we found many cases where the value of $\langle B \rangle$
has nothing to do with the positions of the peaks of $\varphi$. The bad
quality induced by the corresponding error bars is illustrated by 
Fig. \ref{figure5}. Therefore, little trust is available for the energy curve
$e(b)$ that results from this model.

The present work shows that concavity, piecewise at least, is a major
property of any energy surface obtained by a strict minimization  of the
energy under constraint(s). If an energy landscape with ``mountains'' and 
``saddles'' is needed, this concavity contradicts the intuition that energy
transits through local minima. The success of collective models
that use a non-trivial landscape is too strong to be rejected as 
physically and/or mathematically unsound, but its validation likely relates
to other methods such as, resonating group methods \cite{WilTan}, generator 
coordinate (GC) ones \cite{HilGriWhe}, Born-Oppenheimer approximations,
influence functionals \cite{FeyVer}, deconvolutions of wave packets in
collective coordinate spaces \cite {GirGra}, etc. In particular, one might
argue that, while individual exact states $\psi_{\lambda}$ or
mean-field ones $\phi_{\lambda}$ may carry bad error bars for the observable
$b,$ such states may still provide a good global basis for a GC calculation.
But then the physics lies as much in non diagonal elenents $H(b,b')$ and 
$N(b,b')$ of the GC energy and overlap kernels, respectively, than in the
 diagonal, $e(b) \equiv H(b,b).$

Even so, except for anharmonic vibrations, where one valley is sufficient,
one has to justify why convex branches, namely metastability, can be as
important as energy strict minimization.

Because of kinetic energy terms, which enforce delocalizations, a full $H$ is
often not well suited to ensure a good localization of $B$'s, a necessary 
condition for the exploration of an energy surface parametrized by $b$'s.
Recall that, in the Born-Oppenheimer treatment of the hydrogen molecule,
the proton kinetic energy operator is initially removed from the Hamiltonian,
allowing the collective coordinate, namely the interproton distance, to be 
frozen as a zero width parameter. Most often in nuclear, atomic and molecular
physics, such a removal is not available. Hence, ``dangerous consequences''.
This was illustrated by Fig. \ref{figure2}, from the toy Hamiltonian
introduced at the stage of Fig. \ref{figure1}. The growth of fluctuations due
to tunnel effects is spectacular. Fortunately, if one forces the constrained
eigenstate to retain a constant width $\Delta r$ while $\langle r \rangle$
runs, concavity is lost at the profit of a reconstruction of the potential
shape. Note, however, that the reconstruction is not exact; convolution effects
and zero-point kinetic energies still slightly pollute the observed 
``potential''.

We believe that the technical devices used with success in \cite{GLW,FQVVK}
demand a further investigation of the solutions they generated. A hunt for
dangerous, multimodal solutions would make a useful precaution. Since
fluctuations are important at ``phase transitions'', collective
operators have to be completed by their own squares, namely in
combinations of the form, 
$\mathcal{K}=H-\lambda B + \mu(\lambda)  B^2$,
with $\mu(\lambda)$ adjusted to avoid wild variations of $\Delta b$ . Such 
operators $\mathcal{K}$ govern both a constraint and its fluctuation, but 
obviously differ from a double constraint form,
$\mathcal{H}=H-\lambda_1 B+\lambda_2 B^2$, with two independent parameters,
$\lambda_1,\lambda_2$. The second derivative, $d^2 \varepsilon/d\lambda^2$
contains additional terms due to $d\mu/d\lambda$ and $d^2\mu/d\lambda^2$,
hence a one-dimensional path with non concave structures can be induced by
$\mathcal{K}$ inside that concave two-dimensional landscape due to 
$\mathcal{H}$.

In theories using, partly at least, liquid drop models, see \cite{Sierk} for
instance, labels $b$ are purely classical. Such theories are thus safe from
the present warnings. But many other energy surfaces used in realistic
situations come from mean-field constrained calculations. It remains to be
tested whether their solutions, stable or metastable, carry a mechanism that
diminishes the fluctuation of collective degrees of freedom. This mechanism,
if it exists, deserves investigation. We conclude that a review of landscapes
obtained by constrained HF or HB is in order, to analyze the role of
collective coordinate fluctuations. It is clear that such surfaces deserve
corrections because of likely variable widths $\Delta b$ of their collective
observables, and also, obviously, because convolution effects and zero-point
energies must be subtracted.

To summarize, we found both a negative and a positive result.
On the one hand, concavity prevents the emergence of energy landscapes if 
constrained strict minimizations are used with fixed operators $B$ and fixed
trial spaces. Moreover, unacceptable error bars $\Delta b$
can vitiate the meaning of collective labels. On the other hand, given the
same fixed operators and trial spaces, a modest deviation from fixed 
constraints, namely adjustable combinations of $B$ and $B^2,$ commuting 
operators indeed, allows an analysis ``at fluctuations under control'', 
with  unimodal probes of landscapes and a controlled quality of the 
collective representation. A puzzling question is raised: is the 
good quality of constrained mean field solutions in the literature the 
result of a ``self damping'' of collective coordinate fluctuations? Are 
multimodal situations actually blocked when nuclear mass increases?

SK acknowledges support from the National Research Foundation of South Africa.
BG thanks N. Auerbach for stimulating discussions and the University of
Johannesburg  and the Sackler Visiting Chair at the University of Tel Aviv 
for their hospitality during part of this work.

\end{document}